\documentclass[prl,a4paper]{revtex4}
\usepackage{epsfig}

\newcommand{\be}{\begin{equation}}
\newcommand{\ee}{\end{equation}}
\newcommand{\bea}{\begin{eqnarray}}
\newcommand{\eea}{\end{eqnarray}}
\newcommand{\bean}{\begin{eqnarray*}}
\newcommand{\eean}{\end{eqnarray*}}

\newcommand{\gapproxeq}{\lower
.7ex\hbox{$\;\stackrel{\textstyle >}{\sim}\;$}}
\newcommand{\lapproxeq}{\lower
.7ex\hbox{$\;\stackrel{\textstyle <}{\sim}\;$}}

\begin{document}

\bibliographystyle{unsrt}

\title{\bf Glueball hunting in $e^+e^- \to J/\psi\to \phi f_0$}

\author{Frank E. Close$^1$\footnote{e-mail: F.Close1@physics.ox.ac.uk}
and Qiang Zhao$^2$\footnote{e-mail: Qiang.Zhao@surrey.ac.uk}}

\affiliation{1) Department of Theoretical Physics,
University of Oxford, \\
Keble Rd., Oxford, OX1 3NP, United Kingdom}

\affiliation{2) Department of Physics,
University of Surrey, Guildford, GU2 7XH, United Kingdom}

\date{\today}

\begin{abstract}

Building on recent work by Brodsky {\it et al.},
we advocate searching for glueball degrees of freedom 
in $e^+e^- \to J/\psi\to \phi f_0$ at CLEO-c and BES.

\end{abstract}

\maketitle


Brodsky, Goldhaber, and Lee~\cite{brodsky-03} have proposed
a novel approach to producing (scalar) glueballs in $e^+e^-$ annihilation
to account for the anomalously large cross sections for
$J/\psi + \eta_c$, $\chi_{c0}$, and $\eta_c(2S)$ observed
at Belle~\cite{belle}.
They made a pQCD estimate of the cross section for
 $e^+ e^-\to\gamma^*\to J/\psi {\cal G}_0$ at $\sqrt{s}=10.6$ GeV,
and found it to be similar to the
exclusive charmonium-pair production $e^+ e^-\to J/\psi h$
for $h=\eta_c$ and $\chi_{c0}$.
Further, since $\gamma^*\to (c\bar{c})(c\bar{c})$
and $\gamma^*\to (c\bar{c})(gg)$ were of the same nominal order,
they suggested that some portion of the
anomalously large signal observed by Belle in  $e^+ e^-\to J/\psi X$
may actually be due to the production of $J/\psi {\cal G}_0$.

This is an interesting idea theoretically but has a potential limitation
phenomenologically.
As presented, the work of Ref.~\cite{brodsky-03} applies when
$M_{\cal G}\simeq M_{J/\psi} \simeq 3$ GeV.
However, Lattice QCD ~\cite{mp,ukqcd}
and phenomenological studies~\cite{ck,close-farrar-li}
suggest a much smaller mass scale for the lightest scalar glueball
 $M_{\cal G}\simeq 1.5$ GeV.
Some analyses suggest even lower glueball masses~\cite{bugg,bes}.
Taking into account these factors, we anticipate that
the mass scale for the lightest scalar glueball
is smaller than 3 GeV.

Thus we consider here the application of the work of Ref.\cite{brodsky-03} to
the scenario of a scalar glueball in the $O(1)$ GeV mass region.
The analysis of Ref.~\cite{brodsky-03} allows
one to rescale the kinematics such that
instead of a 3-GeV glueball recoiling against
a $J/\psi$, we may consider a 1-GeV glueball recoiling against a $\phi$.
Also, rescaling the c.m.
energy by a factor of three brings one to the kinematic region
of interest currently at BES and to CLEO-c.

In Ref.~\cite{brodsky-03}, the mass scale is introduced
via the mass ratio $r=4m_c/\sqrt{s}$, where $m_c=M_{J/\psi}/2$ is the
charm quark mass. By choosing the glueball mass $M_{\cal G}=M_{J/\psi}$,
the phase space factor for $J/\psi {\cal G}_0$ production cancels
in the branching ratio fraction of $J/\psi {\cal G}_0$ to $J/\psi \eta_c$.
As a result, the $s$-dependence of the branching ratio fraction
will be embedded in $r$ apart from strong couplings and
nonperturbative factors determined through the quarkonium
decay in its rest frame. Due to this feature,
 given that $M_{\cal G}=M_{(q\bar{q})}$,
the branching ratio fraction of
$q_h \bar{q}_h \to\gamma^*\to (q\bar{q}) {\cal G}_0$
to $q_h \bar{q}_h \to\gamma^*\to (q\bar{q}) (q\bar{q})$ scales in terms of $r$
apart from a constant,
where $q_h$ denotes a heavy quark.

\begin{figure}
\begin{center}
\epsfig{file=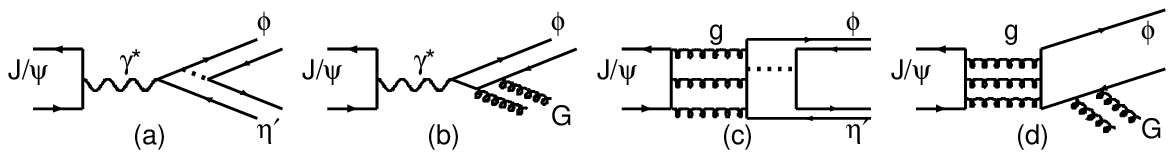, width=16cm,height=6.cm}
\caption{Feynman diagrams for $J/\psi\to \phi \eta^\prime$
and $J/\psi\to \phi {\cal G}_0$
via virtual photon (a)-(b) and three gluon exchanges (c)-(d).
}
\protect\label{fig:(1)}
\end{center}
\end{figure}

First, we examine the process $\gamma^* \to \phi (gg)$
in parallel to $\gamma^* \to J/\psi (gg)$.
An important argument of Ref.~\cite{brodsky-03} is that
the decays of $\gamma^*\to (q\bar{q}) (gg)$ and
$\gamma^*\to (q\bar{q}) (q\bar{q})$ are the same order
(see Fig.~\ref{fig:(1)}(a)-(b)).
The ratio of $\gamma^*\to \phi {\cal G}_0$ to
$\gamma^*\to \mu^+\mu^-$can be estimated
by applying Eq. (7) of Ref.~\cite{brodsky-03}:
\be
\label{eq1}
\frac{R_{\phi{\cal G}_0}}{R_{\mu^+\mu^-}}
=\frac{32 \pi^2\alpha_s^2 e_s^2 r^2(1+r^2/2)\Phi_0^{ee}}{9(1-r^2/4)^2}
\frac{\langle O_1\rangle_\phi}{m_s^3}\frac{|I_0|^2}{s} ,
\ee
where $e_s$ and $m_s=M_\phi/2$ are the $s$ quark's charge
and mass.
The gluon distribution factor $|I_0|^2$ was assumed to be a function of
the glueball's $J^{PC}$ and to scale with mass,
so we adopt the same form as Ref.~\cite{brodsky-03}.
$\Phi_0^{ee}$ is a phase space factor:
$\Phi_0^{ee}=
\sqrt{[1-(M_\phi+M_{\cal G})^2/s][1-(M_\phi-M_{\cal G})^2/s]}$.
Under the condition of $M_{\cal G}=M_\phi$ and $m_s=M_\phi/2$,
we have $\Phi_0^{ee}=\sqrt{1-r^2}$.
For these reduced energies we adopt the running coupling 
constant $\alpha_s \sim 0.33$
at $\sqrt{s}=3.1$ GeV as a guide, and assume also
$\alpha_s^{\cal G}=\alpha_s^{\eta^\prime}=0.33$ in analogy
with the treatment of Ref.~\cite{brodsky-03}.
The matrix element $\langle O_1\rangle_\phi$ is given by the radial
wavefunction of the $s\bar{s}$ in the $\phi$ at the origin $R(0)$ by analogy with the case of $c\bar{c}$:
$\langle O_1\rangle_\phi=|R(0)|^2 N_c/2\pi=2M_\phi f_\phi^2$,
where $f_\phi$ is the decay constant of the $\phi$ meson.

For a glueball mass $M_{\cal G}\simeq 1$ to 1.7 GeV,
by analogy with Ref.~\cite{brodsky-03},
we would compare $\phi{\cal G}$ with $\phi\eta^\prime$ or any of
$\phi f_0(980)$, $\phi f_0(1370)$, $\phi f_0(1500)$, $\phi f_0(1700)$,
which would be clear if $\eta^\prime=\eta(s\bar{s})$
and $f_0=f_0(s\bar{s})$.
However, in practice, the probability of $s\bar{s}$ in $\eta^\prime$ is about
1/2. The scalars are even non-trivial.
The $f_0(980)$ may be a $K\bar{K}$ molecule, or a $q^2\bar{q}^2$
state~\cite{scalar-meson-1,scalar-meson-2}.
In either picture it is not simply related to the
$s\bar{s}$ content of interest to us.
The $f_0(1370)$, $(1500)$, $(1700)$ are believed to be mixtures
of ${\cal G}_0$, $s\bar{s}$ and $n\bar{n}$, so it is not
possible to normalize the $\phi{\cal G}_0$ to these
in a meaningful way~\footnote{Indeed, the $\phi{\cal G}_0$ prediction
will refer in practice to a mixture of these states.
Ultimately, the relative production of these scalars
may help to determine their relative ${\cal G}_0$ contents.}.
Thus we compare $\phi{\cal G}_0$ to $\phi (s\bar{s})$, where
$(s\bar{s})$ is an effective ideal $s\bar{s}$ state with the same mass as $\eta^\prime$.

\begin{figure}
\begin{center}
\epsfig{file=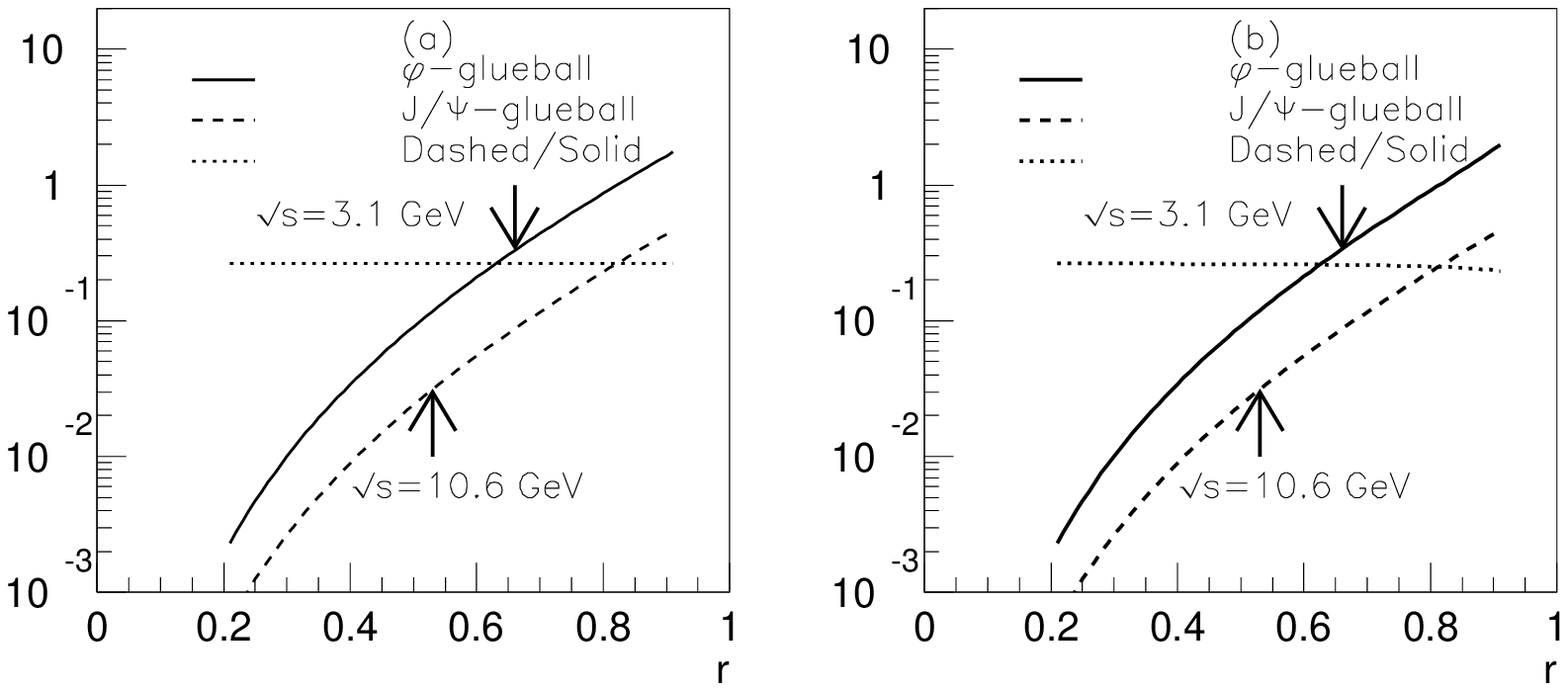, width=12cm,height=7.cm}
\caption{Branching ratio fractions multiplied by $10^4$
for $\phi$-glueball (solid)
and $J/\psi$-glueball (dashed)
production via virtual photons, respectively.
The dotted curve is the ratio of the solid to the dashed, of which the
stable value shows the validity of rescaling the kinematics.
The arrows denote the locations of $r$ corresponding to
the c.m. energies of $\sqrt{s}=3.1$ GeV and 10.6 GeV for these two
reactions.
}
\protect\label{fig:(2)}
\end{center}
\end{figure}

To proceed, the rescaling feature of Eq.~(\ref{eq1})
(i.e. Eq. (7) of Ref.~\cite{brodsky-03}) should be examined.
In Fig.~\ref{fig:(2)}, we present the calculations of the
branching ratio $R_{\phi{\cal G}_0}$
and $R_{J/\psi{\cal G}_0}$ in terms of $r$ to show the rescaling features
between the $\phi$-glueball and $J/\psi$-glueball production
in quarkonium decay via virtual photons.
 The quantity $r$ is in the range of $0<r<1$, which corresponds to the physical
region $\sqrt{s}>4 m_q$.
For the ideal condition that the phase space
factor is cancelled out, the rescaling feature is shown by
the constant fraction (dotted curve in Fig.~\ref{fig:(2)}(a)) between
the $J/\psi$-glueball and $\phi$-glueball production ratios.
The ratio reflects the difference of the
factors ${m_c|I_0|^2}/{\langle O_1\rangle_{\eta_c}}$
and ${m_s|I_0|^2}/{\langle O_1\rangle_\phi}$,
 which denote the ratios of the square of the glueball wavefunctions
at their origins compared to these of the produced quarkonia.
Note that in these two cases the kinematics in terms of $r$ are quite similar
 as indicated by the arrows.
In Fig.~\ref{fig:(2)}(b), we also present the calculation including the
contributions from the non-cancelling phase space factors
with $m_c=1.4$ GeV and $M_{\cal G}=M_{\eta_c}$.
The phase space factor causes
deviations from the exact rescaling between
the solid and dashed curves as $r \to 1$, but is negligible
in most of the kinematical regions.

For glueball mass $M_{\cal G}=M_{\eta^\prime}$,
Eq.~(\ref{eq1}) gives $R_{\phi{\cal G}_0}/R_{\mu^+\mu^-}=9.95\times 10^{-5}(\alpha_s/0.33)^2$.
In association with Eq. (8) of Ref.~\cite{brodsky-03}, 
with $R_{\mu^+\mu^-}=5.88 \ \%$~\cite{pdg2002} we obtain:
\be
\label{eta-prime-em}
br_{J/\psi\to\gamma^*\to \phi{\cal G}}
\simeq br_{J/\psi\to\gamma^*\to \phi (s\bar{s})}=5.85\times 10^{-6}(\alpha_s/0.33)^2
\ee
for the production via virtual photons.
Thus the work of Ref.~\cite{brodsky-03} provides a method
for estimating the virtual photon transitions
in $J/\psi\to \phi (s\bar{s})$, by which the glueball production
can be normalized.
We now examine the consequence of this estimate, and
investigate its prediction for the glueball production.

Apart from the EM transition,
the other important process in $J/\psi\to \phi (s\bar{s})$
is via intermediate gluons,
i.e. $J/\psi\to 3g \to \phi\eta^\prime(s\bar{s})$.
We can thus express the ratio between the EM decay and strong decay
of $J/\psi$ as:
\be
\frac{br_{J/\psi\to 3g\to \phi(s\bar{s})}}
{br_{J/\psi\to \gamma^*\to \phi(s\bar{s})}}
=\frac{br_{J/\psi\to 3g}}{br_{J/\psi\to\gamma^*}}
\frac{br_{\phi\eta^\prime\to 3g}}{br_{\phi\eta^\prime\to \gamma^*}} .
\ee
For an ideal flavor singlet ${\cal F}$,
the ratio for its coupling
to gluons and a virtual photon $\gamma^*$ can be written as
\be
\frac{br_{3g\to {\cal F}}}{br_{\gamma^*\to {\cal F}}}
\sim \frac{\sigma_{\cal F}}{e^2_{\cal F}} ,
\ee
where $\sigma_{\cal F}$ summarises the flavor dependence of the gluon coupling to the 
final state configuration, and $e_{\cal F}$ is the charge
factor of the quarks.
For the ratio of gluon and photon coupling to the initial $J/\psi$ and $s\bar{s}$,
we then have
\be
\frac{br_{J/\psi\to 3g}}{br_{J/\psi\to\gamma^*}}
\frac{br_{\gamma^*\to s\bar{s}}}{br_{3g\to s\bar{s}}}
=\frac{\sigma_{J/\psi}}{e^2_c}\frac{e_s^2}{\sigma_{s\bar{s}}}
\simeq \frac{e_s^2}{e^2_c}=\frac 14 ,
\ee
where we have assumed flavor independence of the quark-gluon coupling. 
With the experimental values,
$br_{J/\psi\to 3g}=0.877 \pm 0.005$ and $br_{J/\psi\to\gamma^*}=0.17 \pm 0.02$~\cite{pdg2002},
we have
\be
\label{ratio-ss}
\frac{br_{3g\to s\bar{s}}}{br_{\gamma^*\to s\bar{s}}}
=4\times\frac{br_{J/\psi\to 3g}}{br_{J/\psi\to\gamma^*}}=21 \pm 3 \ .
\ee
Consequently, we can estimate
\be
\label{strong-br}
br_{J/\psi\to 3g\to \phi (s\bar{s})}
=br_{J/\psi\to \gamma^*\to \phi (s\bar{s})}
\times \frac{br_{J/\psi\to 3g}}{br_{J/\psi\to\gamma^*}}
\frac{br_{3g\to s\bar{s}}}{br_{\gamma^*\to s\bar{s}}}
=(6.5 \pm 1.7)\times 10^{-4} (\alpha_s/0.33)^2 \ ,
\ee
which suggests that
\be
\label{theo}
br^{th}_{J/\psi \to \phi (s\bar{s})}
=br_{J/\psi\to 3g\to \phi (s\bar{s})}
+ br_{J/\psi\to \gamma^*\to \phi(s\bar{s})}
=6.5\times 10^{-4} (\alpha_s/0.33)^2 \ .
\ee

In reality, a pure $s\bar{s}$ state with $J^{PC}=0^{++}$ does not exist: 
the physical scalar states involve
mixing of $s\bar{s}$ with the non-strange $u\bar{u}$ and $d\bar{d}$.
We thus compare $br_{J/\psi\to 3g\to \phi (s\bar{s})}$ with
$br_{J/\psi\to \phi\eta}$ and $br_{J/\psi\to \phi\eta^\prime}$
at the $J/\psi$ mass.
Taking into account the phase space factor, we estimate the
$s\bar{s}$ branching ratio as
\be
br^{exp}_{J/\psi\to \phi (s\bar{s})}
\simeq
br^{exp}_{J/\psi\to \phi\eta}\left(\frac{p_{\eta^\prime}}{p_\eta}\right)^3
+ br^{exp}_{J/\psi\to \phi\eta^\prime}
= (8 \pm 1)\times 10^{-4} \ .
\ee
If we neglect the phase space factor,
the ratio will be $br^{exp}_{J/\psi\to \phi (s\bar{s})}
\simeq (9.8\pm 1.1)\times 10^{-4}$~\cite{pdg2002}, 
which suggests that phase space is not a
significant factor in this estimate.
This comparison shows that Eq.~(\ref{theo}) is in good agreement
with the experimental data
and thereby supports the method of Ref.~\cite{brodsky-03}.
In particular, it provides a way to normalize glueball production
in $J/\psi$ decays.

The above estimate can be applied to $J/\psi\to 3g\to \phi {\cal G}_0$
[Fig.~\ref{fig:(1)} (d)],
which analogous to Eqs.~(\ref{strong-br}) and ~(\ref{theo}) gives
\be
br^{th}_{J/\psi \to \phi {\cal G}_0}
=br_{J/\psi \to 3g \to\phi {\cal G}_0} + br_{J/\psi \to \gamma^* \to\phi {\cal G}_0}
=6.5\times 10^{-4}(\alpha_s/0.33)^2
\ee
in $J/\psi$ decays.

These results implicitly arise because the traditional three-gluon exchange
process is dominant over the EM one in $J/\psi$ decays.
Also, it provides a way to estimate glueball production in $J/\psi$ decays,
which can be normalized by hadron-hadron final states.

\begin{figure}
\begin{center}
\epsfig{file=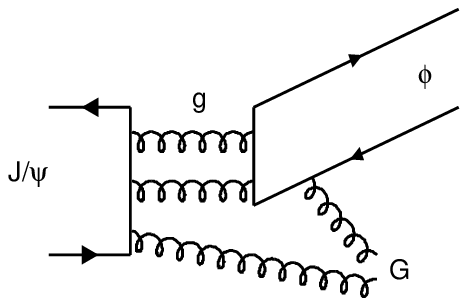, width=7cm,height=7.cm}
\caption{ Diagram for $J/\psi\to \phi {\cal G}_0$
via lower-order gluon exchanges.
}
\protect\label{fig:(3)}
\end{center}
\end{figure}

However, inspecting the gluon exchange process, we note the possible
existence of a lower order diagram for $J/\psi\to \phi{\cal G}_0$,
which could further enhance the glueball production branching ratio.
In Fig.~\ref{fig:(3)}, we show that if the glueball is produced
with one gluon directly from the $c\bar{c}$ annihilation,
its couplng will be $O(1/\alpha_s)$ bigger than
the mechanism of Fig.~\ref{fig:(1)} (c)
assuming all the gluons are perturbative.
We thus estimate the contribution of Fig.~\ref{fig:(3)}:
\be
\label{add}
br^{add}_{J/\psi\to 3g\to \phi{\cal G}_0}
\simeq \frac{1}{\alpha_s^2}br_{J/\psi\to 3g\to \phi{\cal G}_0}
\simeq 6.0\times 10^{-3} .
\ee
We note that the cancellation of the strong coupling constant 
does not mean that $br^{add}_{J/\psi\to 3g\to \phi{\cal G}_0}$
is indepedent of $\alpha_s$. The estimate of Eq.~(\ref{ratio-ss})
should have contained strong coupling $\alpha_s$ in the experimental 
value for $br_{J/\psi\to 3g}/br_{J/\psi\to\gamma^*}$.
With $\alpha_s \sim 0.33$ this enhances the branching ratio to be
\be
\bar{br}^{th}_{J/\psi\to \phi{\cal G}_0}=br^{th}_{J/\psi \to \phi {\cal G}_0}
+br^{add}_{J/\psi\to 3g\to  \phi{\cal G}_0}
\simeq 6.6\times 10^{-3}.
\ee
It shows that if all the gluons are perturbative,
 glueball production would be strongly favored in $J/\psi$ decays and 
suggests that
large glueball production ratios can be
driven by the dominant process of Fig.~\ref{fig:(3)}.
However, cautions should be given to any over-intepretation
of this estimate. We note that the validity of Fig.~\ref{fig:(3)}
dominance will strongly depend on the exchanged gluons being
perturbative, which is not well satisfied as in the case of heavy quark production.
While the actual numbers therefore may be debatable, 
the broad conclusion following from
the Brodsky {\it et al.} approach seem robust.

In summary,
the ideas of Ref.~\cite{brodsky-03} may apply
to $J/\psi\to\phi\eta^\prime$ and in turn to glueball production.
Compared to the subprocess $J/\psi\to \gamma^*\to \phi\eta^\prime$,
scalar glueball production via $J/\psi\to \gamma^*\to \phi {\cal G}_0$
is found to be the same order,
which is consistent with the pQCD calculation
by  Kroll and Passek-Kumer\v{c}ki~\cite{kroll}.
However, for glueball production, it seems likely that
a possible contribution from a lower order diagram may be dominant
over the mechanism of Ref.~\cite{brodsky-03} and the conventional
three-gluon exchange process.
Therefore, we advocate searching for the manifestation 
of glueball degrees of freedom in exclusive
processes, 
$e^+ e^-\to J/\psi \to \phi {\cal G}_0$,
which can be underpinned by the experiments
from the $J/\psi$ factories (BES III, CLEO-c).

We are indebted to D. Bugg for discussions.
This work is supported,
in part, by grants from
the U.K. Particle Physics and
Astronomy Research Council, and the
EU-TMR program ``Eurodice'', HPRN-CT-2002-00311,
and the Engineering and Physical
Sciences Research Council (Grant No. GR/R78633/01).


%



\begin{thebibliography}{99}


%
\bibitem{brodsky-03} S.J. Brodsky, A.S. Goldhaber, and J. Lee,
Phys. Rev. Lett. {\bf 91}, 112001 (2003);
hep-ph/0305269.
%
\bibitem{belle} K. Abe {\it et al.}, [Belle Collaboration],
Phys. Rev. Lett. {\bf 89}, 142001 (2002).
%
\bibitem{mp} C. Morningstar and M. Peardon,
Phys. Rev. D {\bf 56}, 4043 (1997).
%
\bibitem{ukqcd} G. Bali {\it et al.}, UKQCD Collaboration,
Phys. Lett. {\bf B 309}, 378 (1993).
%
%
\bibitem{ck} F.E. Close and A. Kirk,
Phys. Lett. {\bf B 483}, 345 (2000).
%
%
\bibitem{close-farrar-li} F.E. Close, G.R. Farrar, and Z. Li,
        Phys. Rev. D {\bf 55}, 5749 (1997).
%
\bibitem{bugg} D.V. Bugg, M. Peardon, and B.S. Zou,
Phys. Lett. {\bf B 486}, 49 (2000);
P.~Minkowski and W.~Ochs, 
Nucl.\ Phys.\ Proc.\ Suppl.\  {\bf 121}, 123 (2003);
W. Ochs, hep-ph/0311144.
%
\bibitem{bes} BES Collaboration, X.Y. Shen, hep-ex/0209031.
%
\bibitem{scalar-meson-1} F.E. Close and N.A. Tornqvist,
J. Phys. {\bf G 28}, R249 (2002).
%
\bibitem{scalar-meson-2} R.~L.~Jaffe,
Phys.\ Rev.\ D {\bf 15}, 281 (1977).

%
\bibitem{pdg2002} K. Hagiwara {\it et al.} [Particle Data Group],
Phys. Rev. D {\bf 66}, 010001 (2002).
%
\bibitem{kroll} P. Kroll and K. Passek-Kumer\v{c}ki,
Phys. Rev. D {\bf 67}, 054017 (2003).




\end{thebibliography}
\end{document}